\documentclass[10pt,a4paper,twoside]{article}
\usepackage{epsfig}
\usepackage{baltlat6}
\usepackage{array}
\usepackage{here}
\begin{document}
\ \
\vspace{0.5mm}
\setcounter{page}{277}
\vspace{8mm}

\titlehead{Baltic Astronomy, vol.\,XX, XXX--XXX, 2011}

\titleb{INFLUENCE OF MICROLENSING ON SPECTRAL ANOMALY OF LENSED OBJECTS}

\begin{authorl}
\authorb{S. Simi\'c}{1},
\authorb{L.\v C. Popovi\'c}{2}
\authorb{P. Jovanovi\'c}{2}
\end{authorl}

\begin{addressl}
\addressb{1}{Faculty of Science, Department of physics, University of Kragujevac,
\\  Radoja Domanovi\'ca 12, 34000 Kragujevac, Serbia;
ssimic71@gmail.com}
\addressb{2}{Astronomical observatory, Volgina 7,
\\ Belgrade, 11060, Serbia; lpopovic@aob.bg.ac.rs}
\end{addressl}

\begin{summary} Here we consider
the influence of the microlensing on the spectrum of a lensed object taking
into account that composite emission is coming from different regions arranged subsequently
around the central source. We assumed that the lensed object has
three regions with the black body emission; first the innermost with the highest temperature of $10^4K$,
second and third (located around the central) with slightly lower temperatures $7.5\cdot10^3$ and $5\cdot10^3$K, respectively. Than we explore
the flux anomaly in lensed object due to microlensing. We compare U,V and B spectra of a such source.
This  results show that, due to microlensing, in a spectroscopically stratified object a flux anomaly is present.
\end{summary}

\sectionb{1}{INTRODUCTION}

Gravitational macro and microlensing are well known phenomena
that have been widely discussed in the literature
(see e.g., Schneider et al. 1992, Zakharov 1997). The influence of microlensing
on the spectra (continuum and spectral lines) of
lensed QSOs has been investigated theoretically (see e.g. Wambsganss \&
Paczynski 1991; Popovi\'c et al. 2001; Abajas 2002; Abajas et al. 2007, etc.).
Some of the observational effects have also been
presented in several papers (e.g., Lewis et al. 1998; Mediavilla
et al. 1998; Wisotzki et al. 2003;  Richards et al. 2004; Mosquera et al. 2011, Sluse et al. 2011, etc.).
Specially interesting is the flux anomaly observed in a number of quasars (see, e.g. Kratzer et al. 2011).

Microlensing can affect the spectral distribution (SED) of the lensed objects, and it is interesting to explore different situations, i.e. different SEDs and different constellations between the lensed object and (micro)lens.

Here we present  a part of results of flux anomaly investigation in one spectroscopically stratified source that is microlensed by a group of stars.

\sectionb{2}{METHOD}

In order to investigate microlensing influence on lensed object spectra we proceed as follows.
First, we construct an elliptical stratified source containing
three separate regions with different temperatures. The emitting regions are located as a concentric rings
around the central part with diameters mutually differing as an array of integer values starting from 0, taking
that the central part is from 0 to $1d$, next two emitting regions are from $1d$ to $2d$ and from
$2d$ to $3d$, respectively (d is initial diameter of the central part, see Fig. 1, up).

It is assumed that those three emitting regions have different temperature gradients ranging from
the highest temperatures in the central region to lower values in the outer rings. In simulation
we assumed temperature values as $T_1=10^4K, T_2=7.5*10^3K$ and $T_3=5*10^3K$ for the central part, second and third emitting regions, respectively.
Total amount of energy radiated by source regions have a  black-body distribution separately for each specified temperature,
but also depends on the brightness of the surface of each region. It is calculated as:

\begin{equation}
F_i(T_i)=A_i \cdot S_i \cdot B(\nu,T_i)
\label{eq1}
\end{equation}

\noindent where $A_i$ is the brightness of the $i-$th region and $B(\nu,T_i)$ is the black-body spectral distribution. First we assumed that the brightness is the same for  three considered emitting regions. $S_i$ corresponds to the surface of $i-$th emitting region. The summary SED of this stratified object has been obtained as a sum of the three mentioned above emitting regions.

We have calculated magnification map by using the theory presented by Schneider \& Weiss (1987), and apply it separately on each emitting region of the source. Magnification ($m_i$) is calculated as:

\begin{equation}
m_i=\frac{\sum_{k,n} \phi(k,n,T)\ast M(k,n)}{\sum_{k,n}\phi_i(k,n.T)}
\label{eq2}
\end{equation}

\noindent where $\phi$ is the brightness of a pixel on the image of the source and $M(k,n)$ is the matrix of magnification for particular pixels. The sum in denominator and nominator presents the total flux for a particular emitting region with and without microlensing effect, respectively.
Assuming that energy distribution in the source regions has a black body shape, one can express energy radiated from them in the case of a lensed source in a similar way as in Eq. \ref{eq1}, but adding the magnification due to microlensing $m_i$:

\begin{equation}
F^{ml}_i(T_i)=A_i \cdot S_i \cdot B(\nu,T_i) \cdot m_i
\label{eq3}
\end{equation}

Total flux emitted from the lensed source in all wavelength bands is a sum of emission in the three regions, and it is given as:

\begin{equation}
F^{ml}_{tot}=\sum_i F^{ml}_i(T_i)
\label{eq4}
\end{equation}

\begin{equation}
F_{tot}=\sum_i F_i(T_i)
\label{eq5}
\end{equation}

\noindent for the lensed ($F^{ml}_{tot}$) and unlensed ($F_{tot}$) case. This two variables could be compared in order to quantitatively express microlensing influence on the spectrum of quasars.

\sectionb{3}{RESULTS AND DISCUSSION}

We have performed calculation taking that microlensing is caused by a group of 10 Solar mass stars in the lens and get the magnification map in the source plane. The stars are clustered in close space and distributed randomly, while keeping the constant mutual position during the simulation. Also, we created the stratified source as was discussed above which has a central part with diameter of 5 $\mu$as (see Fig. 1 (Up)). For this calculation we assumed that the source is at redshift  $z_l=1$ and lens at $z_s=2.5$.


\begin{figure}[!tH]
\vbox{
\centerline{\psfig{figure=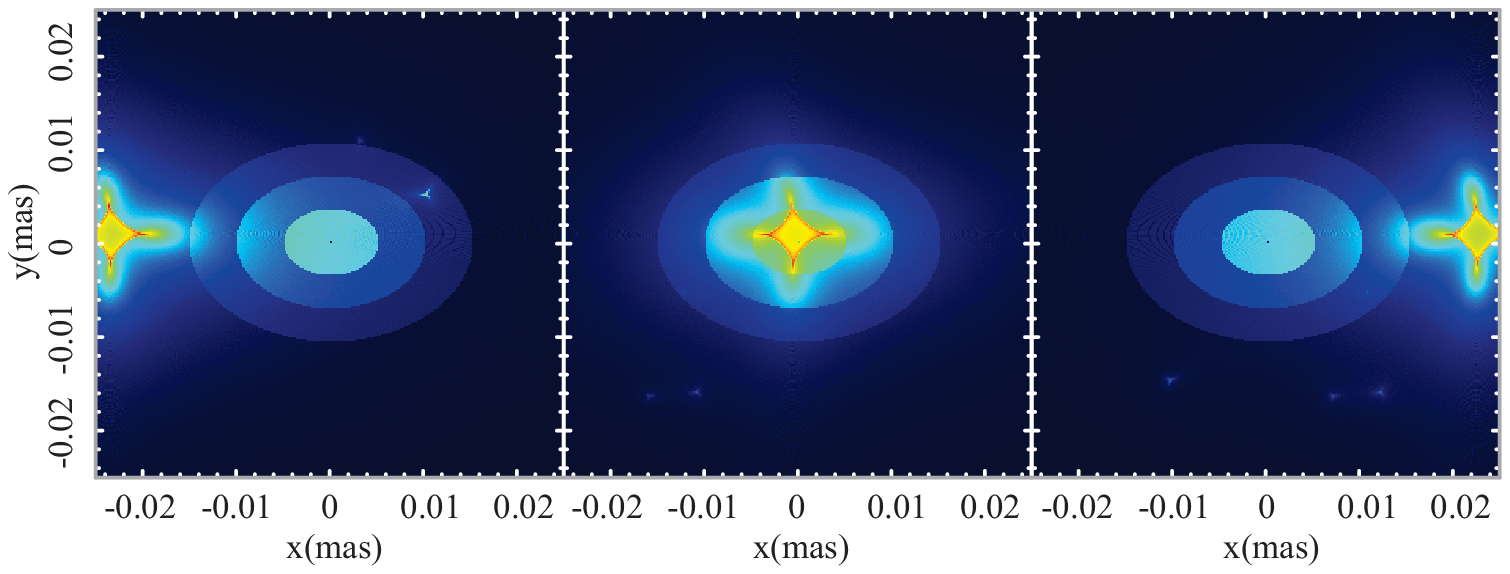,width=88mm,angle=0,clip=}}
\centerline{\psfig{figure=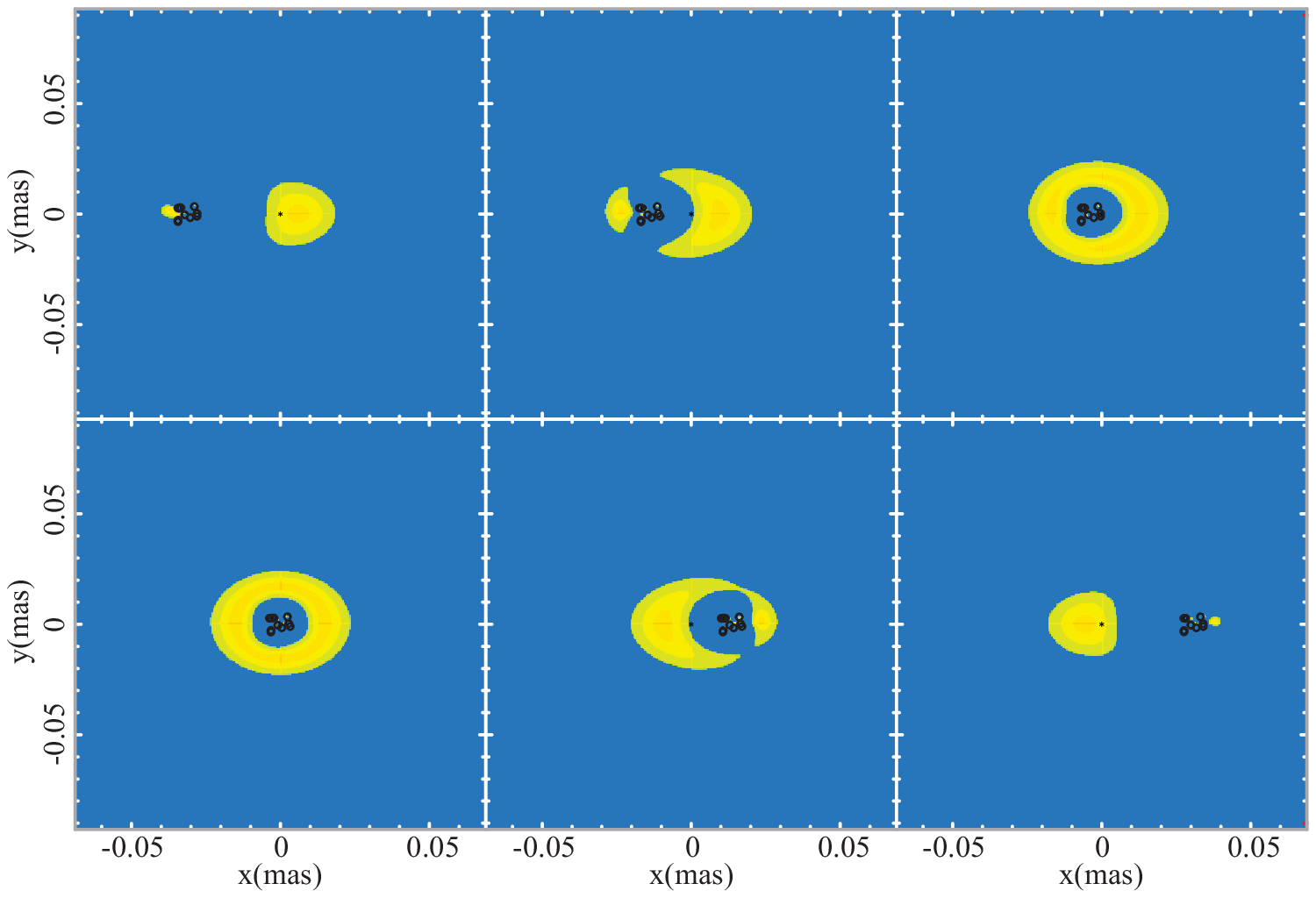,width=88mm,angle=0,clip=}}
\vspace{1mm}
\captionb{1}
{Up: Multi section source and magnification pattern, for three different positions
during the crossing. Down:
Deviation of the lens image due to microlensing of a system of 10 Solar mass stars.}
}
\label{fig01}
\end{figure}

As shown on the images in the Figure 1. (down) microlens has strong influence on the shape and surface intensity of the source. We found that for this, particular situation, the peak of the SED has been shifted approximately 1000 \AA\, that can affect the  observed spectra and cause the flux anomaly. As can be seen in  Fig. 2. the spectrum of the lensed source is increasing in the brightness around two to three times. This could additionally increase possibility for observation of such effect.

The flux ratios between visual $V(5510\pm440){\AA}$, blue $B(4450\pm470){\AA}$ and red $R(6580\pm690){\AA}$ wavelength bands are changing during the microlens event. In Table 1. we presented numerical values of the ratios between $V/B$, $B/R$ and $V/R$ filters for six different positions presented in  Fig. 1 and 2. In this particular case we found that the anomaly in only one image (affected by microlens) the flux anomaly can reach 15\%. The various numerical experiments and detailed discussion of the results will be given elsewhere (Popovic et al. 2011)

\begin{figure}[!tH]
\vbox{
\centerline{\psfig{figure=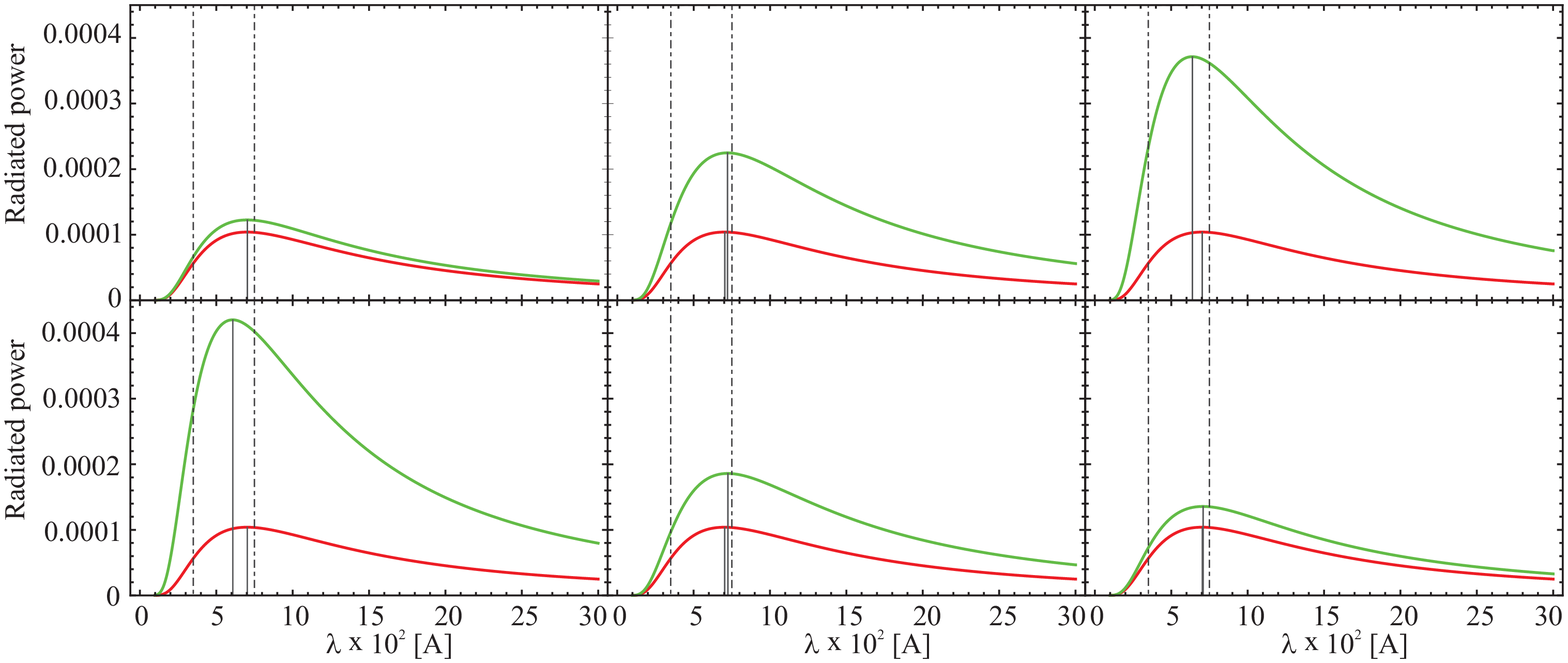,width=100mm,angle=0,clip=}}
\vspace{1mm}
\captionb{2}
{SED changes of the source with stratified spectral emitting regions (see text) modeled for six different positions of microlens shown in Fig. 1.}
}
\label{fig02}
\end{figure}

\begin{table}[!t]
\begin{center}
\vbox{\footnotesize\tabcolsep=3pt
\parbox[c]{124mm}{\baselineskip=10pt
{\smallbf\ \ Table 1.}{\small\
Fractions of emission in V, B and R bands for lens consist of 10 stars randomly distributed in close packed group.\lstrut}}
\begin{tabular}{c|ccc}
\hline
Lens position ERR(mas) & $V/B$ & $B/R$ & $V/R$\hstrut\lstrut\\
\hline
-2066.5 (-0.035) & 1.12273 & 0.535377 & 0.601083 \\
-826.6 (-0.014) & 1.13124 & 0.52803 & 0.597331 \\
-206.65 (-0.0035) & 1.06965 & 0.583469 & 0.624108 \\
0 (0) & 1.04133 & 0.610453 & 0.635685 \\
826.6 (0.014) & 1.13776 & 0.522542 & 0.594525 \\
2066.5 (0.035) & 1.12241 & 0.535558 & 0.601116 \\\hline
\end{tabular}
}
\end{center}
\end{table}

\thanks{This work is a part of the project (176001) "Astrophysical Spectroscopy of
Extragalactic Objects," supported by the Ministry of Science of Serbia.}

\References

\refb Abajas C., Mediavilla E.G., Mu.noz J.A. Popovi\'c L. \v C., \& Oscoz A. 2002, ApJ, 576, 640

\refb Abajas, C., Mediavilla, E., Mu\~noz, J. A. et al. 2007, ApJ, 658, 748

\refb Kratzer, R. M.,  Richards, G.  T.,  Goldberg, D.  M., et al. 2011, ApJL, 728, 18

\refb Lewis G. F., Irwin M. J., Hewett P. C. \& Foltz C. B. 1998, MNRAS, 295, 573

\refb Mediavilla E., Arribas S., del Burgo C., et al. 1998, ApJ, 503, L27

\refb Mosquera, A. M., Mu\~noz, J. A., Mediavilla, E., Kochanek, C. S.  2011, ApJ, 728, 145

\refb Popovi\'c L. \v C., Mediavilla E.G. \& Mu.noz J. 2001, A\&A, 378, 295

\refb Popovi\'c L. \v C., Simi\'c, S., Jovanovi\'c P. et al. 2011, in preparation

\refb Richards G.T., Keeton R.C., Bartosz P. et al. 2004, ApJ, 610, 679

\refb Schneider, P., Ehlers, J., Falco, E. E. 1992, Gravitational Lenses, XIV, 560 pp.
112 figs.. Springer-Verlag Berlin Heidelberg New York.

\refb Schneider P. \& Weiss A. 1987, A\&A, 171, 49.

\refb Sluse, D., Schmidt, R., Courbin, F. et al. 2011, A\&A, 528A, 100

\refb Wambsganss J. \& Paczynski, B. 1991, AJ, 102, 86

\refb Wisotzki L., Becker T., Christensen L., Helms A., Jahnke K., Kelz A., Roth M. M., Sanchez S. F. 2003, A\&A, 408, 455

\refb Zakharov A. F. 1997, Gravitational lenses and microlensing, Janus-K, Moscow.

\end{document}